\documentclass{article}

\begin{document}

\title{An exact plane-fronted wave solution in metric-affine gravity}
\author{Dirk Puetzfeld\thanks{Insitute for Theoretical Physics, University
    of Cologne, D--50923 K\"oln, Germany, E-mail: dp@thp.uni-koeln.de}}
\maketitle

\begin{abstract}
We study plane--fronted electrovacuum waves in metric--affine gravity
(MAG) with cosmological constant in the triplet ansatz sector of the
theory. Their field strengths are, on the gravitational side, curvature
$R_{\alpha}{}^{\beta}$, nonmetricity $Q_{\alpha\beta}$, torsion $T^{\alpha}$
and, on the matter side, the electromagnetic field strength $F$. Here we
basically present, after a short introduction into MAG and its triplet
subcase, the results of earlier joint work with Garc{\'{\i}}a, Mac{\'{\i}}as,
and Socorro \cite{Waves}. Our solution is based on an exact solution of Ozsv\'ath, Robinson,
and R\'ozga describing type N gravitational fields in general relativity as
coupled to electromagnetic null-fields. 

\end{abstract}

\section{Introduction}

Metric-affine gravity (MAG) represents a gauge theoretical
formulation of a theory of gravity which, in contrast to general relativity theory (GR), is no longer confined to a pseudo-Riemannian
spacetime structure (cf.\ \cite{PhysRep}). There are new geometric quantities
emerging in this theory, namely torsion and nonmetricity, which act as
additional field strengths comparable to curvature in the general
relativistic case. Due to this general ansatz, several alternative gravity
theories are included in MAG, the Einstein-Cartan theory e.\thinspace g.,
in which the nonmetricity vanishes and the only surviving post-Riemannian
quantity is given by the torsion. One expects that the MAG provides the
correct description for early stages of the universe, i.\thinspace e.\ at
high energies at which the general relativistic description is expected to break down. In case of vanishing post-Riemannian quantities, MAG proves to be compatible with GR. In contrast to GR, there are presently only a few exact solutions known in MAG (cf.\ \cite{Exact2}),
what could be ascribed to the complexity of this theory.

In the following we will give a short overview of the field equations of
MAG and the geometric quantities featuring therein. Especially, we will present the
results of the work of Obukhov et al.\ \cite{Obukhov} who found that a
special case of MAG, the so called triplet ansatz, is effectively
equivalent to an Einstein-Proca theory. Within the framework of this ansatz,
we will show how one is able to construct a solution of the MAG field
equations on the basis of a plane-fronted wave solution of GR which was
originally presented by Ozsv\'{a}th et al.\ in \cite{Ozsvath}.

\section{MAG in general\label{MAG_KAPITEL}}

In MAG we have the metric $g_{\alpha \beta }$, the coframe $\vartheta
^{\alpha }$, and the connection 1-form $\Gamma _{\alpha }{}^{\beta }$ [with
values in the Lie algebra of the four-dimensional linear group $GL(4,R)$] as
new independent field variables. Here $\alpha ,\beta ,\ldots =0,1,2,3$
denote (anholonomic) frame indices. Spacetime is described by a
metric-affine geometry with the gravitational field strengths nonmetricity $%
Q_{\alpha \beta }:=-Dg_{\alpha \beta }$, torsion $T^{\alpha }:=D\vartheta
^{\alpha }$, and curvature $R_{\alpha }{}^{\beta }:=d\Gamma _{\alpha }{}^{\beta
}-\Gamma _{\alpha }\,^{\gamma }\wedge \Gamma _{\gamma }{}^{\beta }$. A
Lagrangian formalism for a matter field $\Psi $ minimally coupled to the
gravitational potentials $g_{\alpha \beta }$, $\vartheta ^{\alpha }$, $%
\Gamma _{\alpha }{}^{\beta }$ has been set up in \cite{PhysRep}. The
dynamics of this theory is specified by a total Lagrangian 
\begin{equation}
L=V_{\rm{MAG}}(g_{\alpha \beta },\vartheta ^{\alpha },Q_{\alpha \beta
},T^{\alpha },R_{\alpha }{}^{\beta })+L_{\rm{mat}}(g_{\alpha \beta
},\vartheta ^{\alpha },\Psi ,D\Psi ).
\end{equation}
The variation of the action with respect to the independent matter field and the gauge potentials
leads to the field equations: 
\begin{eqnarray}
\frac{\delta L_{\rm{mat}}}{\delta \Psi } &=&0,  \label{matter} \\
DM^{\alpha \beta }-m^{\alpha \beta } &=&\sigma ^{\alpha \beta },
\label{zeroth} \\
DH_{\alpha }-E_{\alpha } &=&\Sigma _{\alpha ,}  \label{first} \\
DH^{\alpha }{}_{\beta }-E^{\alpha }{}_{\beta } &=&\Delta ^{\alpha }{}_{\beta
}.  \label{second}
\end{eqnarray}
Equation (\ref{first}) represents the generalized Einstein equation with the
energy-momentum 3-form $\Sigma _{\alpha }$ as its source whereas (\ref
{zeroth}) and (\ref{second}) are additional field equations which take into
account other aspects of matter, such as spin, shear, and dilation currents
represented collectively by the hypermomentum $\Delta ^{\alpha }{}_{\beta }$. We made use
of the definitions of the gauge field excitations, 
\begin{equation}
M^{\alpha \beta }:=-2\frac{\partial V_{\rm{MAG}}}{ \partial Q_{\alpha \beta }},\quad H_{\alpha }:=-\frac{\partial V_{\rm{MAG}}}{\partial T^{\alpha }},\quad H^{\alpha }{}_{\beta }:=-\frac{\partial V_{\rm{MAG}}}{\partial R_{\alpha
}{}^{\beta }},   \label{exications}
\end{equation}
and of the canonical energy-momentum, the metric stress-energy, and the
hypermomentum current of the gauge fields, 
\begin{equation}
m^{\alpha \beta }:=2\frac{\partial V_{\rm{MAG}}}{\partial g_{\alpha
\beta }},\quad E_{\alpha }:=\frac{\partial V_{\rm{MAG}}}{\partial \vartheta ^{\alpha }},\quad  E^{\alpha }{}_{\beta }:=-\vartheta ^{\alpha }\wedge H_{\beta
}-g_{\beta \gamma }M^{\alpha \gamma }.  \label{gauge_currents}
\end{equation}
Moreover, we introduced the canonical energy-momentum, the metric stress-energy, and the
hypermomentum currents of the matter fields, respectively, 
\begin{equation}
 \sigma ^{\alpha \beta }:=2\frac{\delta L_{\rm{mat}}}{\delta
g_{\alpha \beta }},\quad \Sigma _{\alpha }:=\frac{\delta L_{\rm{mat}}}{\delta \vartheta ^{\alpha }},\quad \Delta ^{\alpha }{}_{\beta }:=\frac{\delta L_{\rm{%
mat}}}{\delta \Gamma _{\alpha }{}^{\beta }}.  \label{matter_currents}
\end{equation}
Provided the matter equation (\ref{matter}) is fulfilled, the following
Noether identities hold: 
\begin{eqnarray}
D\Sigma _{\alpha } &=&\left( e_{\alpha }\rfloor T^{\beta }\right) \wedge
\Sigma _{\beta }-\frac{1}{2}\left( e_{\alpha }\rfloor Q_{\beta \gamma
}\right) \sigma ^{\beta \gamma }+\left( e_{\alpha }\rfloor R_{\beta
}{}^{\gamma }\right) \wedge \Delta ^{\beta }{}_{\gamma },
\label{noether_ident_1} \\
D\Delta ^{\alpha }{}_{\beta } &=&g_{\beta \gamma }\sigma ^{\alpha \gamma
}-\vartheta ^{\alpha }\wedge \Sigma _{\beta }.  \label{noether_ident_2}
\end{eqnarray}
They show that the field equation (\ref{zeroth}) is redundant. Thus we only
need to take into account (\ref{first}) and (\ref{second}). As suggested in \cite{PhysRep}, the most general parity conserving
Lagrangian expressed in terms of the irreducible pieces (cf.\ \cite{PhysRep}) of nonmetricity $Q_{\alpha \beta }$, torsion $T^{\alpha }$%
, and curvature $R_{\alpha \beta }$ reads 
\begin{eqnarray}
V_{\rm{MAG}} &=&\frac{1}{2\kappa } \biggl[ -a_{0}\,R^{\alpha \beta } \wedge \eta
_{\alpha \beta }-2\lambda \eta +T^{\alpha }\wedge \,^{\star}\!\left(
\sum_{I=1}^{3}a_{I}\,^{(I)}T_{\alpha }\right) \nonumber \\
&&+Q_{\alpha \beta }\wedge \,^{\star }\!\left(
\sum_{I=1}^{4}b_{I}\,^{(I)}Q^{\alpha \beta }\right) +b_{5}\left(
^{(3)}Q_{\alpha \gamma }\wedge \vartheta ^{\alpha }\right) \wedge \,^{\star
}\!\left( \,^{(4)}Q^{\beta \gamma }\wedge \vartheta _{\beta }\,\right)  \nonumber \\
&&+2\left( \sum_{I=2}^{4}c_{I}\,^{(I)}Q_{\alpha \beta }\right) \wedge
\vartheta ^{\alpha }\wedge \,^{\star }T^{\beta }\biggr]  \nonumber \\
&&\hspace{-1.57cm}-\frac{1}{2\rho }R^{\alpha \beta }\wedge \,^{\star
}\biggl[\sum_{I=1}^{6}w_{I}\,^{(I)}W_{\alpha \beta
}+\sum_{I=1}^{5}z_{I}\,^{(I)}Z_{\alpha \beta }+w_{7}\,\vartheta _{\alpha
}\wedge \left( e_{\gamma }\rfloor \,^{(5)}W^{\gamma }{}_{\beta }\right) 
\nonumber \\
&&+z_{6}\,\vartheta _{\gamma }\wedge \left( e_{\alpha }\rfloor
\,^{(2)}Z^{\gamma }{}_{\beta }\right) +\sum_{I=7}^{9}z_{I}\,\vartheta
_{\alpha }\wedge \left( e_{\gamma }\rfloor \,^{(I-4)}Z^{\gamma }{}_{\beta
}\right)\biggr].  \label{general_v_mag}
\end{eqnarray}
Note that we decompose the curvature 2-form $R_{\alpha }{}^{\beta }$ into
its antisymmetric and symmetric parts, i.e.\ $R_{\alpha \beta }=W_{\alpha \beta
}+Z_{\alpha \beta }=R_{[\alpha \beta ]}+R_{(\alpha \beta )} \sim$ rotational $\oplus$ strain curvature. The constants entering eq.\ (\ref{general_v_mag}) are the cosmological constant $\lambda $, the weak and strong coupling constant $\kappa $ and $\rho $, respectively, and the 28 dimensionless parameters 
\begin{equation}
a_{0},\dots ,a_{3},b_{1},\dots ,b_{5},c_{2},\dots ,c_{4},w_{1},\dots
,w_{7},z_{1},\dots ,z_{9}.  \label{general_coupling}
\end{equation}
We have here the following dimensions: $[\lambda ]=$length$^{-2}$, $[\kappa ]=$length$^{2}$, $[\rho ]=[\hbar ]=[c]=1.$
The Lagrangian (\ref{general_v_mag}) and the presently known exact solutions in MAG have been
reviewed in \cite{Exact2}.

\section{The triplet ansatz\label{TRIPLET_KAPITEL}}

In the following we will briefly review the results of Obukhov et al.\ \cite
{Obukhov}. Starting from the most general gauge Lagrangian $V_{\rm{MAG}}$ in (\ref{general_v_mag}), we now investigate the special case with 
\begin{equation}
w_{1},\dots ,w_{7}=0,\rm{ \quad }z_{1},\dots ,z_{3},z_{5},\dots
,z_{9}=0,\quad z_{4}\neq 0.  \label{triplet_coupling}
\end{equation}
Thus we consider a general weak part, i.\thinspace e., we do not
impose that one of the weak coupling constants vanishes right from the
beginning. However, the strong gravity part of (\ref{general_v_mag}) is truncated for simplicity. Its only surviving piece is given by$\,$\ the square of the dilation part of the segmental curvature $^{(4)}Z_{\alpha \beta }:=\frac{1}{4}g_{\alpha \beta }Z_{\gamma }{}^{\gamma }$. In this case, the result of Obukhov et al.\ \cite{Obukhov} reads as follows: Effectively, the curvature $R_{\alpha \beta }$ may be considered as Riemannian, torsion and nonmetricity may be represented by a 1-form $\omega
, $
\begin{eqnarray}
Q &=&k_{0}\,\omega ,\quad \quad \Lambda =k_{1}\,\omega ,\quad \quad T=k_{2}\,\omega ,
\label{triplet_allg} \\
T^{\alpha } &=&\,^{(2)}T^{\alpha }=\frac{1}{3}\vartheta ^{\alpha }\wedge T,
\label{torsion_triplet} \\
Q_{\alpha \beta } &=&\,^{(3)}Q_{\alpha \beta }+\,^{(4)}Q_{\alpha \beta }=%
\frac{4}{9}\left( \vartheta _{(\alpha }e_{\beta )}\rfloor \Lambda -\frac{1}{4%
}g_{\alpha \beta }\Lambda \right) +g_{\alpha \beta }Q.
\label{nonmet_triplet}
\end{eqnarray}
With the aid of the Riemannian curvature $\tilde{R}_{\alpha
\beta }$, we denote Riemannian quantities by a
  tilde, the field equation (\ref{first}) looks like the Einstein equation
with an energy-momentum source that depends on torsion and nonmetricity.
Therefore, the field equation (\ref{second}) becomes a system of
differential equations for torsion and nonmetricity alone. In the vacuum
case (i.\thinspace e.\ $\Sigma _{\alpha }=0$ and $\Delta _{\alpha }{}^{\beta
}=0$), these differential equations reduce to 
\begin{eqnarray}
\frac{a_{0}}{2}\,\eta _{\alpha \beta \gamma }\wedge \tilde{R}^{\beta \gamma
}+\lambda \eta _{\alpha } &=&\kappa \Sigma _{\alpha }^{(\omega )},
\label{first_triplet} \\
d\,^{\star }d\omega +m^{2}\,^{\star }\omega &=&0.  \label{second_triplet}
\end{eqnarray}
The four constants $m,$ $k_{0},$ $k_{1},$ and $k_{2}$, which appear in (\ref
{second_triplet}) and (\ref{triplet_allg}), depend uniquely on the
parameters of the MAG Lagrangian (\ref{triplet_coupling}): 
\begin{eqnarray}
k_{0} &=&4\left( a_{2}-2a_{0}\right) \left( b_{3}+\frac{a_{0}}{8}\right)
-3\left( c_{3}+a_{0}\right) ^{2},  \nonumber \\
k_{1} &=&\frac{9}{2}\left( a_{2}-2a_{0}\right) \left( b_{5}-a_{0}\right)
-9\left( c_{3}+a_{0}\right) \left( c_{4}+a_{0}\right) ,  \nonumber \\
k_{2} &=&12\left( b_{3}+\frac{a_{0}}{8}\right) \left( c_{4}+a_{0}\right) -%
\frac{9}{2}\left( b_{5}-a_{0}\right) \left( c_{3}+a_{0}\right) ,  \nonumber \\
m^{2} &=&\frac{1}{z_{4}\kappa }\left( -4b_{4}+\frac{3}{2}a_{0}+\frac{k_{1}}{%
2k_{0}}\left( b_{5}-a_{0}\right) +\frac{k_{2}}{k_{0}}\left(
c_{4}+a_{0}\right) \right) .  \label{effective_coupling_constants}
\end{eqnarray}
The energy-momentum source of torsion and nonmetricity $\Sigma _{\alpha
}^{(\omega )}$, which appears in the effective Einstein equation (\ref
{first_triplet}), reads 
\begin{eqnarray}
\Sigma _{\alpha }^{(\omega )} &=&\frac{z_{4}k_{0}^{2}}{2\rho }\Bigl\{ \left(
e_{\alpha }\rfloor d\omega \right) \wedge \,^{\star }d\omega -\left(
e_{\alpha }\rfloor \,^{\star }d\omega \right) \wedge d\omega \nonumber
\\
&&+m^{2}\left[ \left( e_{\alpha }\rfloor \omega \right) \wedge
\,^{\star }\omega +\left( e_{\alpha }\rfloor \,^{\star }\omega \right)
\wedge \omega \right] \Bigr\}.  \label{triplet_energy_momentum}
\end{eqnarray}
This energy-momentum is exactly that of a Proca 1-form field. 
The parameter $m$ in (\ref{second_triplet}) has the meaning of the mass
parameter ($[m]=$length$^{-1}$). If 
$m$ vanishes, the constrained MAG theory looks similar to the
Einstein-Maxwell theory, as can be seen immediately by comparing (\ref
{triplet_energy_momentum}) with the energy-momentum current of the Maxwell theory 
\begin{equation}
\Sigma _{\alpha }^{\rm{Max}}=\frac{1}{2}\Bigl\{ \left( e_{\alpha }\rfloor
dA\right) \wedge \,^{\star }dA-\left( e_{\alpha }\rfloor \,^{\star
}dA\right) \wedge dA\Bigr\} ,  \label{maxwell_energy_momentum}
\end{equation}
where $A$ denotes the electromagnetic potential 1-form ($F=dA$). Note that
$m=0$ leads to an additional constraint among the coupling constants (cf.\ eq.\ (\ref{effective_coupling_constants})).

\section{Plane-fronted waves in GR\label{PLANE_EINSTEIN_MAXWELL_KAPITEL}}

Ozsv\'{a}th, Robinson and R\'{o}zga \cite{Ozsvath} dealt with a solution
of the Einstein-Maxwell equations. Here we sketch their procedure in order
to show how to generalize it to the triplet subcase of MAG. 
Since we perform our calculations with arbitrary
gravitational coupling constant $\kappa $, the results presented here differ
by some factor of $\kappa $ from the original ones in \cite{Ozsvath}.
Using the coordinates $(\rho ,\sigma ,\zeta ,\bar{\zeta})$, we start with the
line element 
\begin{equation}
ds^{2}=2(\vartheta ^{\hat{0}}\otimes \vartheta ^{\hat{1}}+\vartheta ^{\hat{2}%
}\otimes \vartheta ^{\hat{3}}),  \label{metric}
\end{equation}
and the coframe (the bar denotes complex conjugation) 
\begin{equation}
\vartheta ^{\hat{0}}=\frac{1}{p}\,d\zeta ,\quad \quad \vartheta ^{\hat{1}}=%
\frac{1}{p}\,d\bar{\zeta},\quad \quad \vartheta ^{\hat{2}}=-d\sigma ,\quad
\quad \vartheta ^{\hat{3}}=\left( \frac{q}{p}\right) ^{2}\left( s\,d\sigma
+d\rho \right) .  \label{coframe}
\end{equation}
In order to write down the coframe in a compact form, we made use of the
abbreviations $p$, $q$, and $s$ which are defined in the following way: 
\begin{eqnarray}
p(\zeta ,\bar{\zeta}) &=&1+\frac{\lambda }{6}\,\, \zeta \bar{\zeta},
\label{structure1} \\
q(\sigma ,\zeta ,\bar{\zeta}) &=&\left( 1-\frac{\lambda }{6}\,\,\zeta \bar{\zeta}%
\right) \,\alpha (\sigma )+\zeta \bar{\beta}(\sigma )+\bar{\zeta}\beta
(\sigma ),  \label{strucure2} \\
s(\rho ,\sigma ,\zeta ,\bar{\zeta}) &=&-\frac{\rho ^{2}\lambda }{6}\,\,\alpha
^{2}(\sigma )-\rho ^{2}\beta (\sigma )\,\bar{\beta}(\sigma )+\rho \,\partial
_{\sigma }\left( \ln \,\left| q\right| \right)   \nonumber \\
&&+\frac{p}{2\,q}\,\,H(\sigma ,\zeta ,\bar{\zeta}).  \label{structure3}
\end{eqnarray}
Here $\alpha (\sigma )$, $\beta (\sigma )$, and $H(\sigma ,\zeta ,\bar{\zeta}%
)$ are arbitrary functions of the coordinates and $\lambda $ is the
cosmological constant. Additionally, we introduce the notion of the so-called propagation 1-form $k:=k_{\mu }\vartheta ^{\mu }$ which inherits the
properties of the geodesic, shear-free, expansion-free and twistless null
vector field $k^{\mu }$ representing the propagation vector of a
plane-fronted wave. We proceed by imposing some restrictions on the
electromagnetic 2-form $F$ and a 2-form $S_{\alpha \beta }$ defined in
terms of the irreducible decomposition of the Riemannian curvature 2-form $%
\tilde{R}_{\alpha \beta }$ in the following way:
\begin{eqnarray}
S_{\alpha \beta }&:=&\tilde{R}_{\alpha \beta }-\,^{(6)}\tilde{R}_{\alpha
\beta }=\,^{(1)}\tilde{R}_{\alpha \beta }+\,^{(4)}\tilde{R}_{\alpha \beta } 
\nonumber \\
\, &=&\tilde{R}_{\alpha \beta }+\frac{1}{12}(e_{\nu }\rfloor e_{\mu }\rfloor 
\tilde{R}^{\nu \mu })\,\vartheta _{\alpha }\wedge \vartheta _{\beta }%
\stackrel{\rm{in \,\, vacuum}}{=}\,^{(1)}\tilde{R}_{\alpha \beta }=:C_{\alpha \beta }.  \label{sab}
\end{eqnarray}
They shall obey the so-called radiation conditions 
\begin{equation}
S_{\alpha \beta }\wedge k=0,{\rm \quad and\quad }(e_{\alpha }\rfloor
k)\,S^{\alpha }{}_{\beta }=0,  \label{radiation1}
\end{equation}
and 
\begin{equation}
F\wedge k=0,{\rm \quad and \quad }\frac{1}{2}(e^{\alpha }\rfloor
k)\,e_{\alpha }\rfloor F=0.  \label{radiation2}
\end{equation}
If one imposes the conditions (\ref{radiation1}) and (\ref{radiation2}), the formerly arbitrary functions $\alpha (\sigma)$ and $\beta (\sigma )$ in (\ref{structure1})-(\ref{structure3}) become restricted, namely $\alpha(\sigma)$ to the real and $\beta(\sigma)$ to the complex domain, see \cite{Ozsvath}.
In the next step we insert the ansatz for the coframe (\ref{coframe}) into
the Einstein-Maxwell field equations with cosmological constant 
\begin{eqnarray}
\eta _{\alpha \beta \gamma }\wedge \tilde{R}^{\beta \gamma }+2\,\lambda
\,\eta _{\alpha } &=&2\kappa \Sigma _{\alpha }^{\rm{Max}},
\label{einsteineq} \\
dF=0,\quad d\,\,^{\star }F &=&0,  \label{maxwelleq2}
\end{eqnarray}
where $\Sigma _{\alpha }^{\rm{Max}}$ is the energy-momentum current of the
electromagnetic field (cf.\ eq.\ (\ref{maxwell_energy_momentum})).
Let us consider the vacuum field equations first before switching on the
electromagnetic field, i.\thinspace e.\ there are only gravitational waves and
we only have to take into account the left hand side of (\ref{einsteineq}). 
As shown in \cite{Ozsvath}, this equation, after inserting
the coframe into it (cf.\ eq.\ (4.34) of \cite{Ozsvath}), turns into a homogeneous
PDE for the unknown function $H(\sigma ,\zeta ,\bar{\zeta})$:
\begin{equation}
H_{,\zeta \bar{\zeta}}+\frac{\lambda }{3p^{2}}H=0.  \label{vacuumPDE}
\end{equation}
Equation (4.39) of \cite{Ozsvath} supplies us with the solution for $H(\sigma ,\zeta ,\bar{\zeta})$ in terms of an arbitrary holomorphic function $\phi (\sigma ,\zeta )$ 
\begin{equation}
H(\sigma ,\zeta ,\bar{\zeta})=\phi _{,\zeta }-\frac{\lambda }{3}\frac{\bar{%
\zeta}}{p}\,\phi +\bar{\phi}_{,\bar{\zeta}}-\frac{\lambda }{3}\frac{\zeta }{p}%
\,\bar{\phi}.  \label{homogeneSol}
\end{equation}
Observe that $H$ in (\ref{homogeneSol}) is a real quantity. We are now going
to switch on the electromagnetic field. We make the following ansatz for the
electromagnetic 2-form $F\,\ $in terms of an arbitrary complex function $f(\sigma ,\zeta )$:
\begin{equation}
F=dA=-d\left[ \left( \int^{\zeta }d\zeta ^{^{\prime }}\,f(\zeta ^{^{\prime
}},\sigma )+\int^{\bar{\zeta}}d\bar{\zeta}^{^{\prime }}\bar{f}(\bar{\zeta}%
^{^{\prime }},\sigma )\right) \vartheta ^{\hat{2}}\right] .
\label{faradansatz}
\end{equation}
In compliance with \cite{Ozsvath}, this ansatz for $F$ leads to $\Sigma _{\alpha }^{\rm{Max}}=-2\,\delta _{\alpha }^{\hat{2}}\,p^{2}f\bar{f}\,\vartheta ^{\hat{0}}\wedge \vartheta ^{\hat{1}}\wedge \vartheta ^{\hat{2}}$.
Now the field equations (\ref{einsteineq})-(\ref{maxwelleq2}) turn into an inhomogeneous PDE for $H(\sigma ,\zeta ,\bar{\zeta})$ (cf.\ eq.\ (4.35) of \cite{Ozsvath}): 
\begin{equation}
H_{,\zeta \bar{\zeta}}+\frac{\lambda }{3p^{2}}H=\frac{2\kappa p}{q}\,f\,\bar{%
f}.  \label{inhomogenePDE}
\end{equation}
The homogeneous solution $H_{\rm{h}}(\sigma ,\zeta ,\bar{\zeta})$ of this
equation is again given by (\ref{homogeneSol}).
The particular solution $H_{\rm{p}%
}(\sigma ,\zeta ,\bar{\zeta})$ of the inhomogeneous equation can be written
in a similar form, 
\begin{equation}
H_{\rm{p}}(\sigma ,\zeta ,\bar{\zeta})=\mu _{,\zeta }-\frac{\lambda }{3}%
\frac{\bar{\zeta}}{p}\mu +\bar{\mu}_{,\bar{\zeta}}-\frac{\lambda }{3}\frac{%
\zeta }{p}\bar{\mu},  \label{particularSOL}
\end{equation}
where the function $\mu (\sigma ,\zeta ,\bar{\zeta})$ can be expressed in
the following integral form: 
\begin{equation}
\mu (\sigma ,\zeta ,\bar{\zeta})=\kappa \int^{\bar{\zeta}}d\bar{\zeta}%
p^{2}\int^{\zeta }d\zeta ^{^{\prime }}\frac{1}{p^{2}}\int^{\zeta ^{^{\prime
}}}d\zeta ^{^{\prime \prime }}\frac{p\,f\,\bar{f}}{q}.  \label{mue_part}
\end{equation}
Of course, one is only able to derive $H_{\rm{p}}$ explicitly after
choosing the arbitrary functions $\alpha (\sigma )$ and $\beta (\sigma )$ in
(\ref{structure1})-(\ref{structure3}). They enter the coframe and, as
a consequence, the function $\mu (\sigma ,\zeta ,\bar{\zeta})$ in (\ref
{mue_part}). The general solution reads\footnote{We changed the name of the vacuum solution from $H(\sigma ,\zeta ,\bar{\zeta})$, mentioned in (\ref{homogeneSol}), into $H_{\rm{h}}(\sigma,\zeta ,\bar{\zeta}).$} 
\begin{equation}
H(\sigma ,\zeta ,\bar{\zeta})=H_{\rm{h}}(\sigma ,\zeta ,\bar{\zeta})+H_{%
\rm{p}}(\sigma ,\zeta ,\bar{\zeta}).
\label{alllg_loesung_homogen_plus_partikulaer}
\end{equation}
We proceed with a particular choice for the functions entering the coframe and
the ansatz for the electromagnetic potential
\begin{equation}
\alpha =1,\quad \quad \beta =0,\quad \quad f=f_{0}\,\zeta ^{n}\rm{\quad
where\quad }n=0,\pm 1,\pm 2,\dots \,,  \label{alphabetaf}
\end{equation}
and {$[f_{0}]=$length$^{-2-n}$. The electromagnetic potential is now given by
\begin{equation}
A=-f_{0}\left( \int^{\zeta }d\zeta ^{^{\prime }}\,\zeta ^{^{\prime }n}+\int^{%
\bar{\zeta}}d\bar{\zeta}^{^{\prime }}\bar{\zeta}^{^{\prime }n}\right)
\vartheta ^{\hat{2}}.  \label{specpotential}
\end{equation}
The only unknown function is $H_{\rm{p}}(\sigma ,\zeta ,\bar{\zeta})$.
Thus we have to carry out the integration (cf.\ (7.7) of \cite{Ozsvath}) in
eq.\ (\ref{mue_part}), after substituting the function $f=f_{0}\,\zeta ^{n}$
which originates from our ansatz (\ref{alphabetaf}). The solution of
this integration for different choices of $n$ is given in (7.10)-(7.13) of 
\cite{Ozsvath}. We will present this solution in a more compact form
as\bigskip\newline
\underline{(i) $n<-1$} 
\begin{eqnarray}
H_{\rm{p}} &=&\frac{2\kappa pf_{0}^{2}}{q}\left( \frac{(\zeta \bar{\zeta}%
)^{1+n}}{(1+n)^{2}}+4\left( \frac{\lambda }{6}\right) ^{-n-1}\ln \left|
\,q\,\right| -4\left( \frac{\lambda }{6}\right) ^{-n-1}\ln \left|
\,p-1\,\right| \right.  \nonumber \\
&&\left. +4\sum_{r=1}^{-n-1}\frac{\left( \frac{\lambda }{6}\right) ^{-n-r-1}%
}{r\,\left( \zeta \bar{\zeta}\right) ^{r}}\right) +\frac{8\kappa
f_{0}^{2}\left( \zeta \bar{\zeta}\right) ^{n+1}}{(1+n)\,p},
\label{hpkleinerSOL}
\end{eqnarray}
\underline{(ii) $n=-1$} 
\begin{equation}
H_{\rm{p}}=\frac{2\kappa f_{0}^{2}}{p}\left( 4\,q\,\ln \left| \,q\,\right|
+\frac{2\lambda \zeta \bar{\zeta}}{3}\ln \left( f_{0}^{2}\zeta \bar{\zeta}%
\right) +\frac{q}{2}\ln ^{2}\left( f_{0}^{2}\zeta \bar{\zeta}\right) \right)
,  \label{hpgleichSOL}
\end{equation}
\underline{(iii) $n>-1$ } 
\begin{eqnarray}
H_{\rm{p}} &=&\frac{8\kappa qf_{0}^{2}}{p}\left( \frac{\lambda }{6}\right)
^{-n-1}\left( \ln \left| \,q\,\right| +\sum_{r=1}^{n}\frac{\left(
_{r}^{n}\right) }{r}\left( \left( p-2\right) ^{r}-(-1)^{r}\right) \right) 
\nonumber \\
&&+\frac{2\kappa f_{0}^{2}(\zeta \bar{\zeta})^{n+1}}{p\,(n+1)^{2}}(4(n+1)+q).
\label{hpgroesserSOL}
\end{eqnarray}
Note that the original solution in \cite{Ozsvath} is not completely correct, as
admitted by Ozsv\'{a}th \cite{PrivComm}. Needless to say that we displayed in (\ref{hpkleinerSOL})-(\ref{hpgroesserSOL}) the correct expressions for $H_{\rm{p}}$.
We characterize the solutions obtained above by means of the selfdual part
of the conformal curvature 2-form $^{+}C_{\alpha \beta },$ the trace-free
Ricci 1-form 
$\tilde{R}_{\alpha }\!\!\!\!\!\!\!\!\!\nearrow $, and the Ricci scalar $\tilde{R}$. Here we list only the results for the
general ansatz (\ref{structure1})-(\ref{structure3}), i.\thinspace e.\ for
arbitrary $\alpha (\sigma )$, $\beta (\sigma )$, and $H(\sigma ,\zeta ,\bar{%
\zeta})$. For the sake of brevity we make use of the structure functions $p$
and $q$ as defined in (\ref{structure1}) and (\ref{strucure2}): 
\begin{eqnarray}
^{+}C_{\hat{2}\hat{0}} &=&-\,^{+}C_{\hat{0}\hat{2}}=\frac{1-i}{4}\,\partial
_{\zeta }\left[ q^{2}\partial _{\zeta }\left( \frac{p}{q}H\right) \right]
\,\vartheta ^{\hat{0}}\wedge \vartheta ^{\hat{2}}, \\
^{+}C_{\hat{2}\hat{1}} &=&-\,^{+}C_{\hat{1}\hat{2}}\,=\frac{1+i}{4}\,\partial
_{\bar{\zeta}}\left[ q^{2}\partial _{\bar{\zeta}}\left( \frac{p}{q}H\right) %
\right] \,\vartheta ^{\hat{1}}\wedge \vartheta ^{\hat{2}}, \\
\tilde{R}_{\hat{2}}\!\!\!\!\!\!\!\!\!\nearrow\,\,\,\,%
&=&\frac{2pq\kappa p}{q}\,f\,\bar{f}\,\vartheta ^{\hat{2}}=2\kappa p^{2}\,f\,%
\bar{f}\,\vartheta ^{\hat{2}}.
\end{eqnarray}

\section{Plane-fronted waves in MAG\label{PLANE_MAG_KAPITEL}}

We now turn to the triplet subcase of MAG. Thus we are concerned with the
triplet of 1-forms in eq.\ (\ref{triplet_allg}).
As shown in section \ref{TRIPLET_KAPITEL}, the triplet ansatz reduces the electrovacuum MAG field equations (\ref{first}) and (\ref{second}) to an effective Einstein-Proca-Maxwell system: 
\begin{eqnarray}
a_{0}\,\eta _{\alpha \beta \gamma }\wedge \tilde{R}^{\beta \gamma }+2\lambda
\eta _{\alpha } &=&2\kappa \left[ \Sigma _{\alpha }^{(\omega )}+\Sigma
_{\alpha }^{\rm{Max}}\right] ,  \label{triplet_feq_1} \\
d\,^{\star }d\omega +m^{2}\,^{\star }\omega &=&0,  \label{triplet_feq_2} \\
dF=0,\quad d\,^{\star }F &=&0.  \label{triplet_feq_4}
\end{eqnarray}
From here on we will presuppose that $m^{2}=0$, reducing eq.\ (\ref{triplet_feq_2}) to $d\,^{\star }d\omega =0$ and the energy-momentum $\Sigma_{\alpha }^{(\omega )}$ of the triplet field to the first line of eq.\ (\ref{triplet_energy_momentum}).
As one realizes immediately, the system (\ref{triplet_feq_1})-(\ref
{triplet_feq_4}) now becomes very similar to the one investigated in the
Einstein-Maxwell case in (\ref{einsteineq})-(\ref{maxwelleq2}). 
Let us start with the same ansatz for the line element, coframe (\ref{metric})-(\ref{structure3}), and electromagnetic 2-form (\ref{faradansatz}). The only thing
missing up to now is a suitable ansatz for the 1-form $\omega $ which
governs the non-Riemannian parts of the system and enters eqs.\ (\ref
{triplet_feq_1})-(\ref{triplet_feq_2}): 
\begin{equation}
\omega =-\left[ \int^{\zeta }d\zeta ^{^{\prime }}g(\sigma ,\zeta ^{^{\prime
}})+\int^{\bar{\zeta}}d\bar{\zeta}^{^{\prime }}\bar{g}(\sigma ,\bar{\zeta}%
^{^{\prime }})\right] \vartheta ^{\hat{2}}.  \label{ansatz_omega}
\end{equation}
Here $g(\sigma ,\zeta )$ represents an arbitrary complex function of the
coordinates. 
Since the first field equation (\ref{triplet_feq_1}) in the MAG case differs from the Einsteinian one only by the emergence of $\Sigma _{\alpha}^{(\omega )}$. Accordingly, we expect only a linear change in the PDE (\ref{inhomogenePDE}). Thus, in
case of switching on the electromagnetic and the triplet field, the solution for ${\cal H}(\sigma ,\zeta ,\bar{\zeta})$ entering the
coframe, is determined by 
\begin{equation}
{\cal H}_{,\zeta \bar{\zeta}}+\frac{\lambda }{3p^{2}}{\cal H}=\frac{%
2\kappa p}{q}\left( f\bar{f}+g\bar{g}\right) .  \label{inhom_PDE_triplet}
\end{equation}
We use calligraphic letters for quantities that belong to the MAG
solution. Consequently, the homogeneous solution ${\cal H}_{\rm{h}}$, corresponding
to $f=g=0$, is given again by (\ref{homogeneSol}).
In order to solve
the inhomogeneous equation (\ref{inhom_PDE_triplet}), we modify the ansatz
for $H_{\rm{p}}$ made in (\ref{particularSOL}) and (\ref{mue_part}). For
clarity, we distinguish between the Einstein-Maxwell and the MAG case by
changing the name of $\mu (\sigma ,\zeta ,\bar{\zeta})$ in (\ref
{particularSOL}) into $M(\sigma ,\zeta ,\bar{\zeta})$ which leads to the
following form of ${\cal H}_{\rm{p}}$: 
\begin{equation}
{\cal H}_{\rm{p}}(\sigma ,\zeta ,\bar{\zeta})=M_{,\zeta }-\frac{\lambda 
}{3}\frac{\bar{\zeta}}{p}M+\bar{M}_{,\bar{\zeta}}-\frac{\lambda }{3}\frac{%
\zeta }{p}\bar{M},  \label{H_p_MAG}
\end{equation}
\begin{equation}
{\rm where} \quad M=\kappa \int^{\bar{\zeta}}d\bar{\zeta}p^{2}\int^{\zeta }d\zeta ^{^{\prime }}%
\frac{1}{p^{2}}\int^{\zeta ^{^{\prime }}}d\zeta ^{^{\prime \prime }}\frac{p\,%
}{q}\left( f\,\bar{f}+g\bar{g}\right) .  \label{M_MAG}
\end{equation}
Thus, the general solution of (\ref{inhom_PDE_triplet}) is given by 
\begin{equation}
{\cal H}(\sigma ,\zeta ,\bar{\zeta})={\cal H}_{\rm{h}}(\sigma ,\zeta
,\bar{\zeta})+{\cal H}_{\rm{p}}(\sigma ,\zeta ,\bar{\zeta})\rm{.}
\label{general_h_MAG}
\end{equation}
Substitution of this ansatz into the field equations yields the following
constraint for the coupling constants of the constrained MAG Lagrangian: 
\begin{equation}
a_{0}=1,\quad z_{4}=\frac{\rho }{2k_{0}}.
\label{plane_relation_among_coupling_constants}
\end{equation}
Here, we made use of the definition of $k_{0}$ mentioned in (\ref{effective_coupling_constants}). We will now look for a particular solution ${\cal H}_{\rm{%
p}}$ of (\ref{inhom_PDE_triplet}). As in the Riemannian case, we choose
$\alpha =1$, $\beta =0$ and make a polynomial ansatz for the functions $f$ and
$g$ which govern the Maxwell and triplet regime of the system, 
\begin{equation}
f=f_{0}\zeta ^{n}\quad n=0,\pm 1,\pm 2,\dots \,,\quad g =g_{0}\zeta ^{l}\quad l=0,\pm 1,\pm 2,\dots \,\,\,,  \label{ansatz_g_MAG}
\end{equation}
with $[g_{0}]=$length$^{-2-l}$. Now we have to perform the integration in (\ref
{M_MAG}) which yields the solution for ${\cal H}_{\rm{p}}$ via eq.\ (\ref
{H_p_MAG}). At this point we remember that the solution for ${\cal H}_{%
\rm{p}}$, in case of excitations corresponding to $f$, is already known
from eqs.\ (\ref{hpkleinerSOL})-(\ref{hpgroesserSOL}). Let us introduce a new name for $H_{\rm{p}}$ as displayed in (\ref
{hpkleinerSOL})-(\ref{hpgroesserSOL}), namely ${\cal H}_{\rm{p}}^{f}$.
Furthermore, we introduce the quantity ${\cal H}_{\rm{p}}^{g}$, defined
in the same way as ${\cal H}_{\rm{p}}^{f}$ but with the ansatz for $g$ 
from (\ref{ansatz_g_MAG}). Thus, one has to perform the substitutions 
$n\rightarrow l$ and $f_{0}\rightarrow g_{0}$ in eqs.\ (\ref{hpkleinerSOL})-(\ref{hpgroesserSOL}) in order to obtain ${\cal H}_{\rm{p}}^{g}$. Due to 
the linearity of our ansatz in (\ref{M_MAG}), we can infer
that the particular solution ${\cal H}_{\rm{p}}$ in the MAG case is
given by the sum of the appropriate branches for ${\cal H}_{\rm{p}}^{f}$
and ${\cal H}_{\rm{p}}^{g}$, i.e.\ the general solution form eq.\ (\ref
{general_h_MAG}) now reads  
\begin{equation}
{\cal H}(\sigma ,\zeta ,\bar{\zeta})={\cal H}_{\rm{h}}(\sigma ,\zeta ,\bar{\zeta})+{\cal H}_{\rm{p}%
}^{f}(\sigma ,\zeta ,\bar{\zeta})+{\cal H}_{\rm{p}}^{g}(\sigma ,\zeta ,%
\bar{\zeta}).  \label{sol_h_MAG_aufgespalten}
\end{equation}
Note that we have to impose the same additional constraints among the
coupling constants as in the general case (cf.\ eq.\ (\ref
{plane_relation_among_coupling_constants})).
In contrast to the general relativistic case, there are two new geometric
quantities entering our description, namely the torsion $T_{\alpha }$ and the
nonmetricity $Q_{\alpha \beta }$ given by
\begin{eqnarray}
Q_{\alpha \beta }&=&-\frac{4k_{1}}{9}\vartheta _{(\alpha }e_{\beta )}\rfloor \left[
\int^{\zeta }d\zeta ^{^{\prime }}g(\sigma ,\zeta ^{^{\prime }})+\int^{\bar{%
\zeta}}d\bar{\zeta}^{^{\prime }}\bar{g}(\sigma ,\bar{\zeta}^{^{\prime }})%
\right] \vartheta ^{\hat{2}}  \nonumber \\
&&+g_{\alpha \beta }\left( \frac{k_{1}}{9}-k_{0}\right) \left[ \int^{\zeta
}d\zeta ^{^{\prime }}g(\sigma ,\zeta ^{^{\prime }})+\int^{\bar{\zeta}}d\bar{%
\zeta}^{^{\prime }}\bar{g}(\sigma ,\bar{\zeta}^{^{\prime }})\right]
\vartheta ^{\hat{2}}, \\
T^{\alpha } &=&-\frac{k_{2}}{3}\left[ \int^{\zeta }d\zeta ^{^{\prime }}g(\sigma ,\zeta ^{^{\prime
}})+\int^{\bar{\zeta}}d\bar{\zeta}^{^{\prime }}\bar{g}(\sigma ,\bar{\zeta}%
^{^{\prime }})\right] \vartheta ^{\alpha }\wedge \vartheta ^{\hat{2}}.
\end{eqnarray}

\section{Summary}

We investigated plane-fronted electrovacuum waves in MAG with cosmological
constant in the triplet ansatz sector of the theory. The spacetime under
consideration carries curvature, nonmetricity, torsion, and an
electromagnetic field. Apart from the cosmological constant, the solution
contains several arbitrary functions, namely $\alpha (\sigma )$, $\beta
(\sigma )$, $f(\sigma ,\zeta )$, $g(\sigma ,\zeta )$, and $\phi (\sigma
,\zeta )$. One may address these functions by the generic term \textit{wave
parameters} since they control the different sectors of the solution like
the electromagnetic or the non-Riemannian regime. In this way, we
generalized the class of solutions obtained by Ozsv\'{a}th, Robinson, and
R\'{o}zga \cite{Ozsvath}. 
\newline\newline
The author is grateful to Prof. F.W. Hehl, C. Heinicke, and G. Rubilar
for their help. The support of our German-Mexican collaboration by CONACYT--DFG (E130--655---444 MEX 100) is gratefully acknowledged.

\end{document}